\documentclass[epsf,usegraphicx]{mn2e}

\usepackage{xspace}

\title[RAT J1953+1859 -- a new dwarf nova] 
{RAT J1953+1859: a dwarf nova discovered through high amplitude 
QPOs in quiescence}

\author[]
{Gavin Ramsay$^{1}$, Pasi Hakala$^{2}$, Thomas Barclay$^{1,3}$, 
Peter Wheatley$^{4}$, George Marshall$^{4}$, \and Harry Lehto$^{2}$,
Ralf Napiwotzki$^{5}$, Gijs Nelemans$^{6}$, Stephen Potter$^{7}$, 
Ian Todd$^{8}$\\
$^{1}$Armagh Observatory, College Hill, Armagh, BT61 9DG\\
$^{2}$Tuorla Observatory, University of Turku, V\"ais\"al\"antie 20, FIN-21500
Piikki\"o, Finland\\
$^{3}$Mullard Space Science Laboratory, University College London,
Holmbury St. Mary, Dorking, Surrey, RH5 6NT\\
$^{4}$Department of Physics, University of Warwick, Coventry, CV4 7AL\\
$^{5}$Centre for Astrophysics Research, STRI, University of Hertfordshire, 
Hatfield, AL10 9AB\\
$^{6}$Department of Astrophysics, IMAPP, Radboud University Nijmegen, 
P.O. Box 9010, NL-6500 GL, Nijmegen, The Netherlands\\
$^{7}$South African Astronomical Observatory, P.O. Box 9, Observatory
7935, Cape Town, South Africa\\
$^{8}$Astrophysics Research Centre, Queen's University, Belfast, BT7 1NN\\
}

\begin{document}
\outer\def\gtae {$\buildrel {\lower3pt\hbox{$>$}} \over 
{\lower2pt\hbox{$\sim$}} $}
\outer\def\ltae {$\buildrel {\lower3pt\hbox{$<$}} \over 
{\lower2pt\hbox{$\sim$}} $}
\newcommand{\ergscm} {ergs s$^{-1}$ cm$^{-2}$}
\newcommand{\ergss} {ergs s$^{-1}$}
\newcommand{\ergsd} {ergs s$^{-1}$ $d^{2}_{100}$}
\newcommand{\pcmsq} {cm$^{-2}$}
\newcommand{\ros} {\sl ROSAT}
\newcommand{\chan} {\sl Chandra}
\newcommand{\xmm} {\sl XMM-Newton}
\def\rchi{{${\chi}_{\nu}^{2}$}}
\newcommand{\Msun} {$M_{\odot}$}
\newcommand{\Mwd} {$M_{wd}$}
\def\Mdot{\hbox{$\dot M$}}
\def\mdot{\hbox{$\dot m$}}
\newcommand{\teff}{\ensuremath{T_{\mathrm{eff}}}\xspace}
\newcommand{\src} {RAT J1953+1859}

\maketitle

\begin{abstract}

We report the discovery of an accreting binary, {\src}, made during
the RApid Temporal Survey (RATS) on the Isaac Newton Telescope. It
showed high amplitude (0.3 mag) quasi-periodic oscillations on a
timescale of $\sim$20 mins. Further observations made using the Nordic
Optical Telescope showed it to be $\sim$4 mag brighter than in the
discovery images. These photometric observations, together with radial
velocity data taken using the William Herschel Telescope, point to an
orbital period of $\sim$90 mins. These data suggest that {\src} is a
dwarf novae of the SU UMa type.  What makes {\src} unusual is that it
is the first such system to be discovered as a result of high
amplitude QPOs during quiescence. This suggests that high-cadence
wide-field surveys could be another means to discover cataclysmic
variables as a result of their short period variability.

\end{abstract}


\section{Introduction}

The RApid Temporal Survey (RATS) is a deep, wide-field, fast-cadence
photometric survey which allows the discovery of objects which vary in
their intensity on timescales ranging from a few minutes to several
hours (Ramsay \& Hakala 2005). Whilst our primary aim is to discover
AM CVn binaries (the hydrogen deficient cataclysmic variables, CVs,
with orbital periods less than $\sim$70 min) we expect to discover
objects ranging from contact binaries, pulsating stars, flare stars
and accreting binaries in general.  In our pilot survey, we discovered
more than 40 new variable objects including one system which we
identified as a short period pulsating sdB star (Ramsay et al 2006).

Since our pilot survey, we have obtained further wide-field camera
data using the 2.5m Isaac Newton Telescope (INT) on La
Palma. Currently, we have discovered many 1000's of new variable
stars. Most of them have periods longer than a few hours, but a number
were found to show periods shorter than 30 mins (and hence candidate
AM CVn binaries). For these systems we have obtained followup
spectroscopy. Here we report on observations on one such source, RAT
J1953+1859, which is an accreting binary system.

\section{Wide Field Camera discovery data}

Data were taken using the Wide Field Camera on the 2.5m INT on 30 May
2005. There are 4 CCDs with approximate total field of view of
33$^{'}\times33^{'}$ (with a 11$^{'}\times11^{'}$ gap missing on one
of the corners). The field was centered on the globular cluster M71
which has a tidal radius of 9.0$^{'}$ (Harris 1996). A series of 30
sec white light exposures were taken for 100 min -- this includes the
$\sim$30 sec readout time of the camera. Data were flat fielded and
bias subtracted in the usual manner.

Variable sources were identified using the difference image analysis
package {\tt Dandia} (see Bramich et al 2005 for a description).  Full
details of how we identified variable sources will be given elsewhere.
Suffice to say, we identified a strongly variable object which was
modulated on a period of $\sim$20 min with an amplitude of 0.3
mag. The light curve and the corresponding power spectrum are shown in
Figure 1. While the dominant period is at 19.7 min there are also
prominent peaks at $\sim$15.5 and $\sim$9.2 min. We pre-whitened the
light curve on a period of 19.7 min period, and found that the peaks
at 15.5 and 9.2 min were still present. We therefore do not believe
these peaks are related to the window function. We phased the data on
all three periods and find that they are not strictly periodic, rather
they are quasi-periodic oscillations (QPOs).

To determine the sky co-ordinates of this variable source, we
identified objects in the field which were in the 2MASS catalogue. We
then used {\tt astrom} (Wallace \& Gray 2002) to obtain the
astrometric solution for the field.  The position of the variable
source is $\alpha$=19$^{h}$ 53$^{m}$ 27.2$^{s}$, $\delta$= +18$^{o}$
59$^{'}$ 13.5$^{''}$ (2000) and the residuals on the positions are
0.4$^{''}$. This places it 13.3$^{'}$ distant (equating to 1.5
$\times$ the cluster tidal radius) from the globular cluster M71. We
therefore consider it unlikely that the variable source is associated
with the cluster. We show the finding chart in Figure \ref{chart}.

\begin{figure}
\begin{center}
\setlength{\unitlength}{1cm}
\begin{picture}(8,12)
\put(-0.5,-1.5){\includegraphics{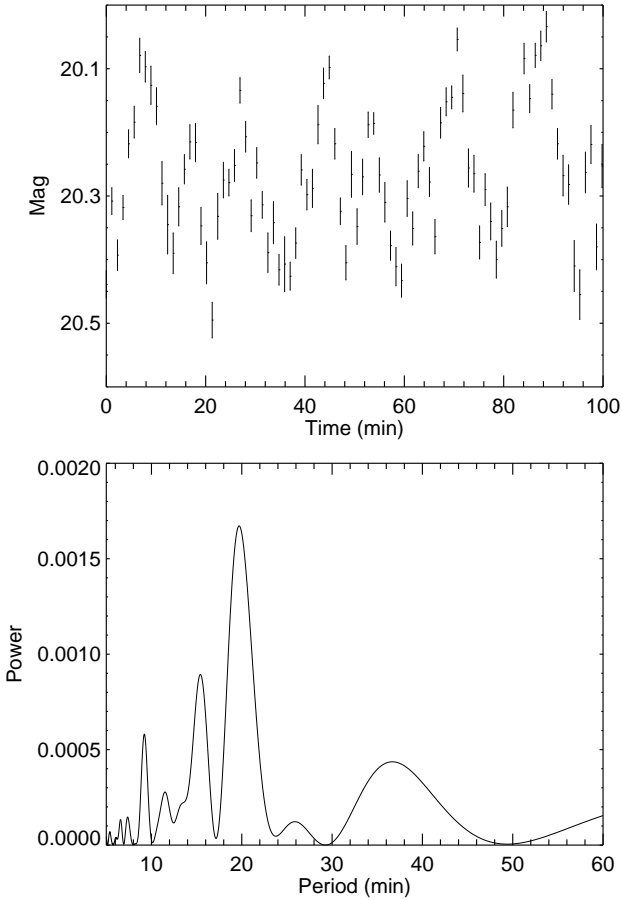}}
\end{picture}
\end{center}
\caption{Upper panel: The light curve of RAT J1953+1859 taken using
the Wide Field Camera on the INT in June 2005. The data were taken in
white light and each exposure was 30 sec. Lower panel: The power spectrum
of the light curve shown in the upper panel.}
\label{light}
\end{figure}

We took $BVI$ images prior to the sequence of white light
exposures. Although we did not obtain images of photometric standard
fields we were able to place our filter data on the standard system by
matching up objects which were in the catalogue of Geffert \& Maintz
(2000) who obtained photometry of stars in the field of M71.  Our
variable source was $V\sim$20.4 and $(B-V)\sim$0.2 at the time of our
observations (we note that our $BV$ data was not
simultaneous). Compared with other stars in the same field, our
variable source is clearly blue.

\begin{figure*}
\begin{center}
\setlength{\unitlength}{1cm}
\begin{picture}(8,7)
\put(-4,0){\includegraphics{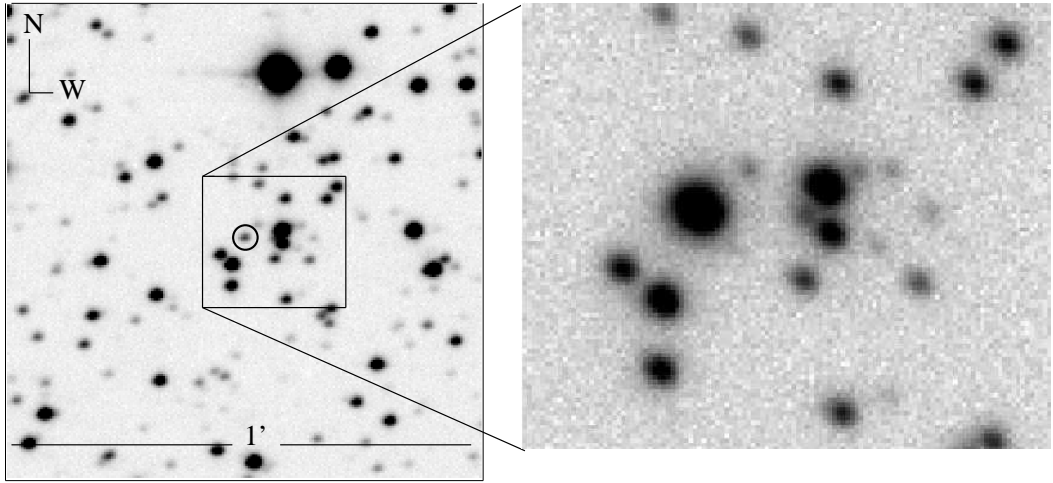}}
\end{picture}
\end{center}
\caption{Left hand panel: The white light finding chart for RAT
J1953+1859 made using the Wide Field Camera on the INT in June
2005. The circle shows the position of RAT J1953+1859. Right hand
panel: The image taken the NOT on Sept 28th 2008 -- RAT J1953+1859 was
$\sim$4 mag brighter than in the INT image.}
\label{chart}
\end{figure*}

\section{Followup photometry}
\label{not}

We obtained further photometry of RAT J1953+1859 using the 2.5m Nordic
Optical Telescope (NOT) sited on La Palma on 28th Sept 2008 using
ALFOSC.  It was immediately clear that RAT J1953+1859 was much
brighter than in the discovery data (the right hand panel of Figure
\ref{chart}). We did not obtain any filtered data of the field, but
comparison with our INT white light images suggest that {\src} was
$V\sim$16.5, or $\sim$4 mag brighter than our INT discovery data.

We proceeded to obtain a sequence of 15 sec exposures in white light
which lasted for 140 min.  The chip was windowed to reduce readout
time to 5 sec.  We show the full light curve in the top left hand
panel of Figure \ref{not}.  There is some evidence that the light
curve repeats itself after $\sim$90 mins with an amplitude of
$\sim$0.4 mag, although we note that the observation length was 140
mins. The second highest peak in the power spectrum is at $\sim$46
mins (lower left hand panel of Figure \ref{not}). We removed the
$\sim$90 min trend and the resulting `residual' light curve is shown
in the top right hand panel of Figure \ref{not}. It is clear even by
`eye' that low amplitude ($\sim$0.02 mag) quasi-periodic behaviour is
seen in the first half of the light curve.  The power spectrum of this
residual light curve is shown in the lower right hand panel of Figure
\ref{not} and shows peaks near 5.7, 10.2 and 12.7 mins.  None of these
peaks coincides with the peaks seen in the power spectra of the data
taken using the INT.

\begin{figure*}
\begin{center}
\setlength{\unitlength}{1cm}
\begin{picture}(14,10)
\put(-3,-12){\includegraphics{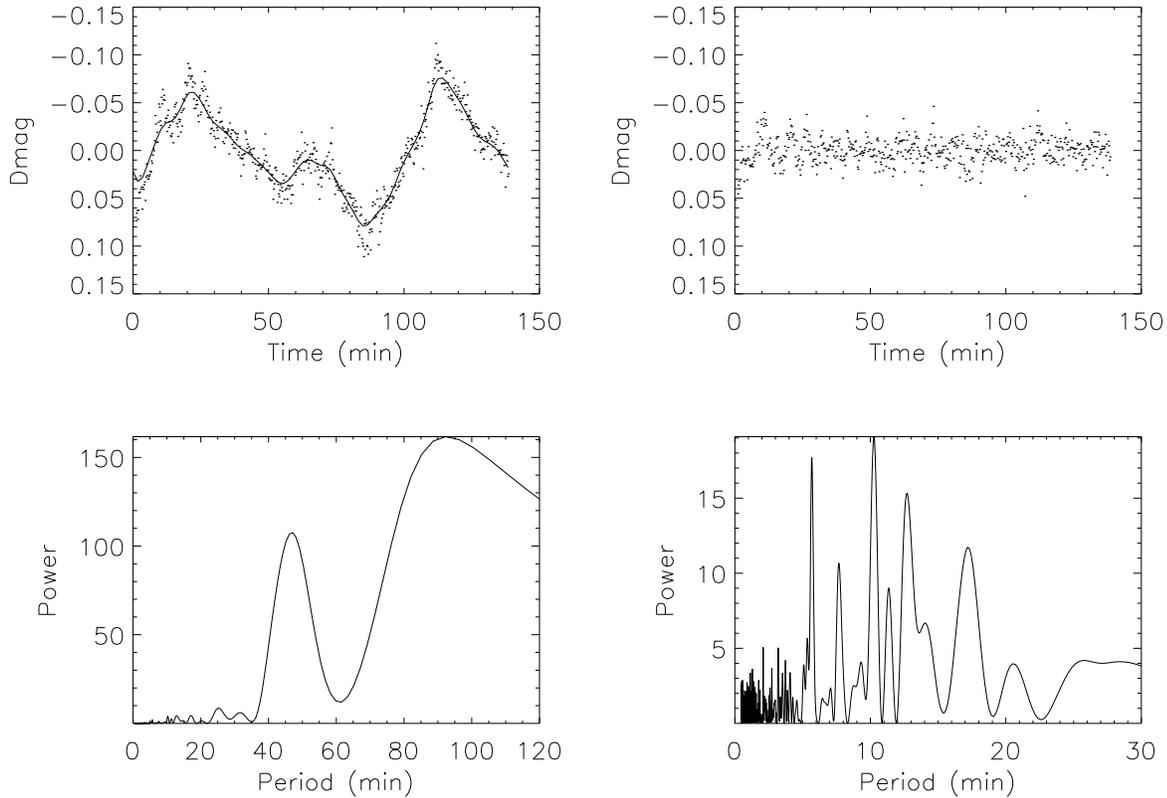}}
\end{picture}
\end{center}
\caption{Photometric observations made using the Nordic Optical
Telescope on 28th Sept 2008. In the top left panel we show our full
light curve and its power spectrum in the lower left.  In the top
right hand panel we show the light curve with the general trend
removed. Clear QPO like features are seen in the light curve with
periods between 5 and 13 min (lower right hand panel).}
\label{not}
\end{figure*}

\section{Optical Spectroscopy}
\label{spec}

\subsection{Main spectral features}

We obtained spectra of RAT J1953+1859 using the 4.2m William Herschel
Telescope (WHT) and the Intermediate dispersion Spectrograph and
Imaging System (ISIS) on La Palma at three separate epochs (Table 1).
All the data were bias subtracted and the spectra were created using
optimal extraction. Since we only took one arc lamp observation at the
start and end of each sequence we cross-correlated the sky spectra and
applied this small correction (less than 1 pixel) to the spectra.

\begin{figure*}
\begin{center}
\setlength{\unitlength}{1cm}
\begin{picture}(16,8.5)
\put(0.5,-0.5){\includegraphics{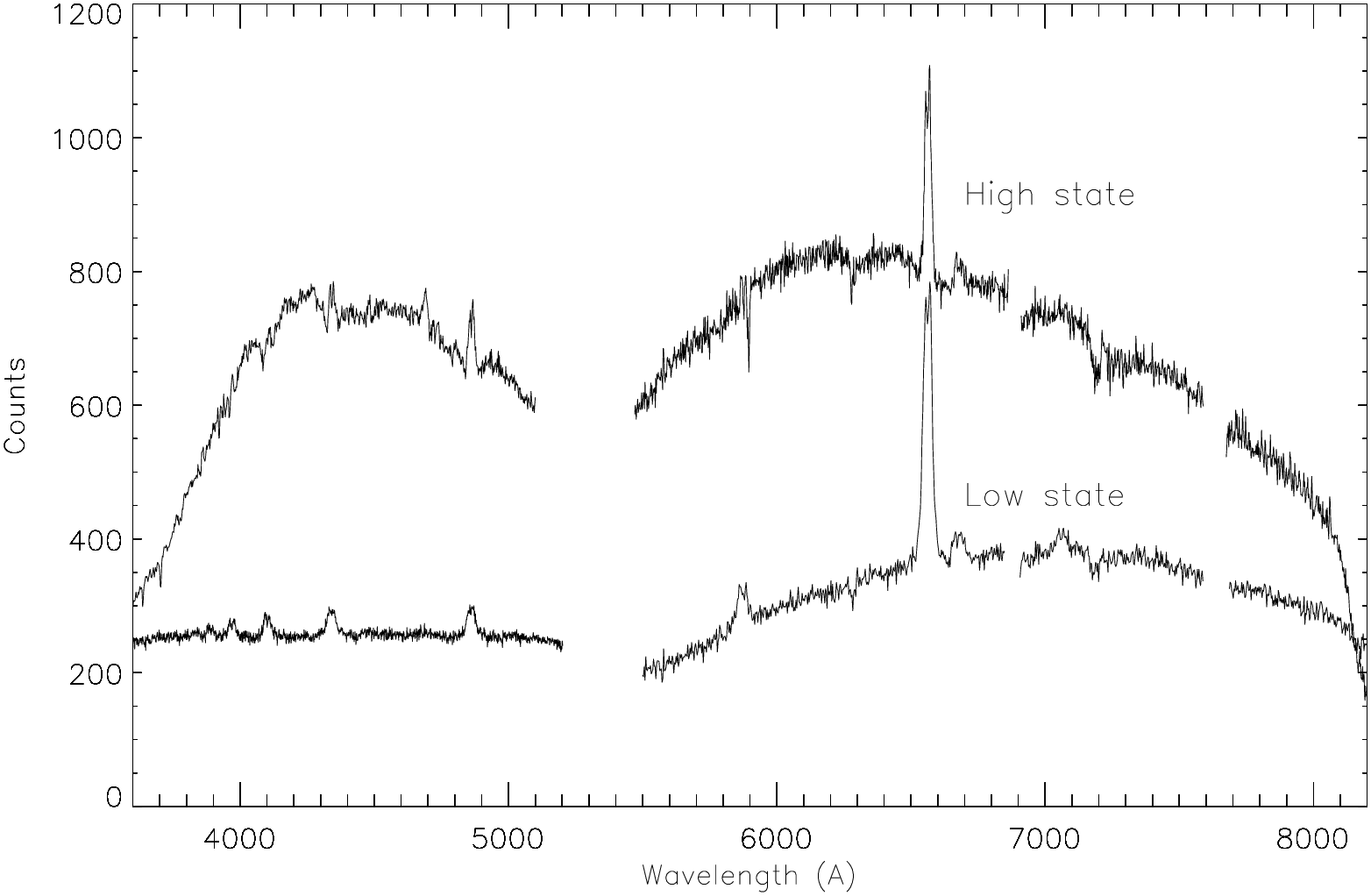}}
\end{picture}
\end{center}
\caption{The blue and red spectra of RAT J1953+1859 taken using ISIS
on the WHT on Sep 29 2008 and Oct 6 2008 when the source was in an outburst
and quiescence respectively. We have not flux corrected the spectra so 
the instrumental response is partly reflected in the data.}
\label{spec-low-high}
\end{figure*}

Our first set of spectra, which had exposures ranging from 120 sec to
420 sec and were taken when the source was in quiescence, shows
emission lines of H$\alpha$, H$\beta$ and H$\gamma$ decreasing in
prominence (Figure \ref{spec-low-high}). The Balmer lines are also
split (most clearly in H$\alpha$) and broad (a FWHM of $\sim$40\AA,
corresponding to velocities of 1800 km/s) indicating the presence of
an accretion disk. We also note the presence of a He I emission line
at 5876 and 6678 \AA \hspace{1mm} but the absence of the He II 4686
\AA line.  Strong Balmer emission is typical of dwarf novae in
quiescence, although the He II 4686 \AA \hspace{1mm} line is present
in some dwarf novae (eg YZ Cnc, Shafter \& Hessman 1988), but not in
others (eg SS Cyg Martinez-Pais et al 1994).

The second epoch observations were taken on the same night as we
obtained observations using the NOT when RAT J1953+1859 was in
outburst. The exposure time was 90 sec with a further 12 sec for
readout. By the epoch of the third observation, the source had
returned to quiescence and the exposure time was 240 sec with a
further 25 sec for readout. We show the mean of the spectra taken in
the blue and red arms in Figure \ref{spec-low-high}. The spectra again
show H$\alpha$ strongly in emission with the subsequent Balmer lines
decreasing in strength with some evidence for the emission lines lying
in a broader absorption core. At H$\delta$ there is no sign of
emission. In contrast to the quiescent state spectrum, He II (4686
\AA) is seen as a weak emission line. These spectral features
are typical of a dwarf nova in outburst (see Martinez-Pais et al 1996
for spectral observations of SS Cyg taken over an outburst).

\begin{table}
\begin{center}
\caption{The log of our optical spectral observations of RAT J1953+1859
made using the WHT and ISIS. We show the date of the observations, the 
number of spectra in each arm, the 
gratings used (red and blue respectively), the slit width, the 
spectral resolution in the red and blue arms and the accretion state 
of the source (QU - quiescence, OB - outburst).}
\begin{tabular}{lrcrrr}
\hline
Date & No. & Gratings & Slit  & Res & Accn \\
     & Spec   & & $^{''}$ &     & State \\
\hline
2008 Aug 03 & 6 & R158R R158B & 0.6 & 5\AA, 5\AA & QU \\
2008 Sep 29 & 24 & R316R R300B & 1.5 & 8\AA, 5\AA & OB \\
2008 Oct 06 & 53 & R158R R300B & 0.8 & 8\AA, 3\AA & QU\\
\hline
\end{tabular}
\end{center}
\label{spec-log}
\end{table}

\subsection{A search for periods}

\begin{figure}
\begin{center}
\setlength{\unitlength}{1cm}
\begin{picture}(16,6.5)
\put(0.,-0.5){\includegraphics{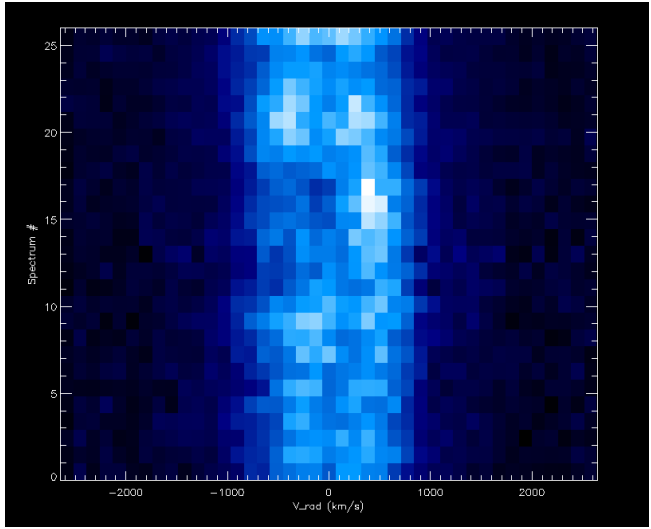}}
\end{picture}
\end{center}
\caption{The stacked spectra showing the H$\alpha$ emission line taken
on Oct 6 2008 when the source was in a quiescent state. Each spectrum
has been normalised so the continuum is unity and time increases
vertically.}
\label{spec-low-stack}
\end{figure}

We show the spectra centered on H$\alpha$ in Figure
\ref{spec-low-stack}. Although there are clear variations in the
spectral profile over time, there are no obvious repeating features.
We then performed a search for periods by determining the radial
velocity of the H$\alpha$ emission features by fitting one or more
Gaussian components -- some spectra can be well fitted using a single
Gaussian, but in others three components were required. We tried a
number of approaches, including fixing the line width of the higher
velocity components. However, we were not able to find convincing
evidence for any periods in the data.

\begin{figure*}
\begin{center}
\setlength{\unitlength}{1cm}
\begin{picture}(16,10)
\put(0,0){\includegraphics{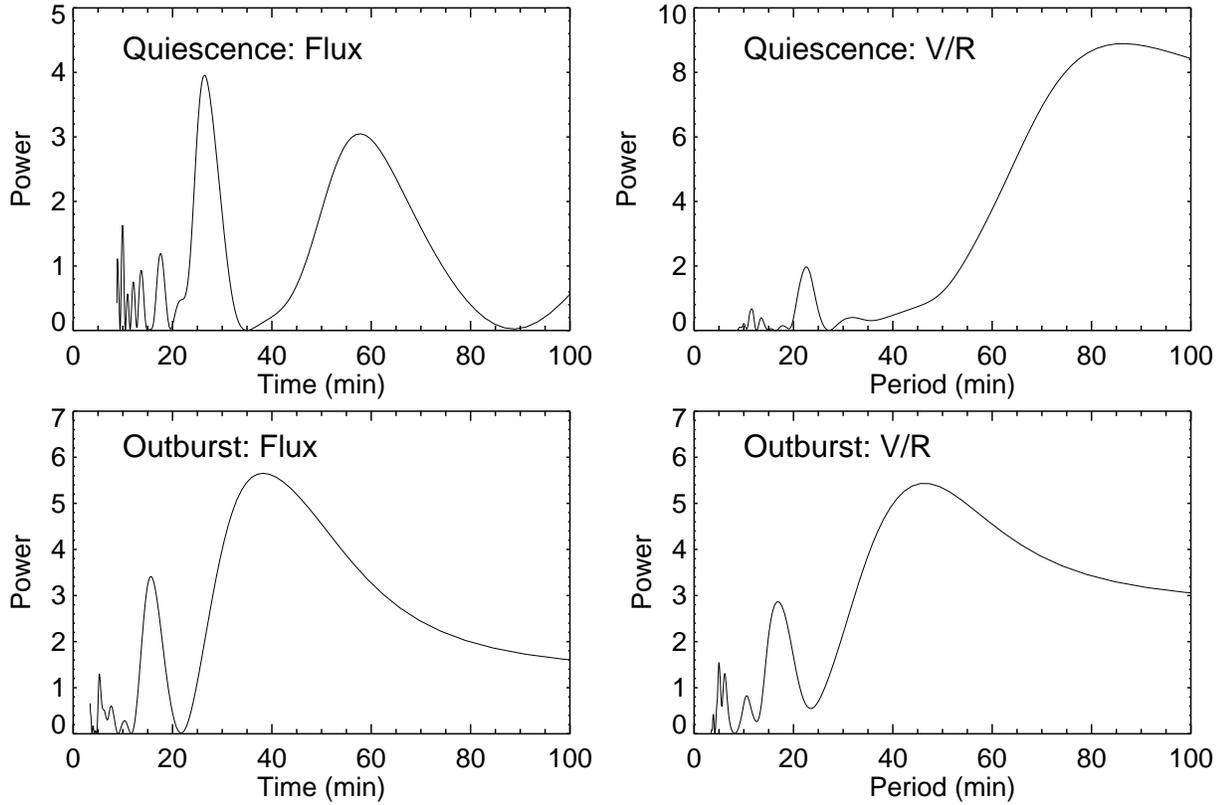}}
\end{picture}
\end{center}
\caption{The top panel shows the power spectra of the flux (left hand
panel) and the V/R ratio (right hand panel) of the H$\alpha$ line in
quiescence. In the lower panel we show the same plots made using data
taken in outburst (in Oct 2008).}
\label{v-over-r}
\end{figure*}

Our second approach has the minimum of assumptions and splits the
emission line into `violet' and `red' components (the `V/R' ratio) at
an arbitrary, but fixed, wavelength.  We then performed a Lomb-Scargle
power spectrum on the total flux and the resulting V/R time
series. The longer time series made on Oct 6th 2008 when the source
was in quiescence, gives a peak in the flux power spectra near $\sim$1
hr and 27 min, and peaks near 80--90 min and 22 min in the V/R ratio
power spectra (Figure \ref{v-over-r}). The shorter time series made
when the source was in outburst shows a peak near 40 mins and 16 mins
in both the flux and V/R time series data.

\section{{\sl ROSAT} observations}

A pointed observation of the field of RAT J1953+1859 was made on
10th Apr 1994 using the {\sl ROSAT} High Resolution Imager (pass band
0.2--2.4keV) which lasted for 31.7 ksec. Extracting an image using
this data shows a clear X-ray source close to the position of RAT
J1953+1859. Although not obvious from this image, there are two
sources noted in the {\sl ROSAT} HRI pointed catalogue which are
4.1$^{''}$ and 6.2$^{''}$ distant from the optical position of the
variable object and have quoted count rates of 0.0034(6) and 0.0057(8)
counts/sec. In the subsequent analysis we assume that the wavelet
analysis which was used in making the HRI catalogue has mistakenly
distinguished two sources when there is only one in reality. We
therefore believe that we have identified the X-ray counterpart of RAT
J1953+1859 and has a count rate of 0.0091 ct/s in the HRI.

We took the event data and corrected the arrival times of every event
to the solar system barycenter. We searched for a periodic modulation
in the X-ray light curve using a Discrete Fourier Transform. There is
no evidence for any modulation period in the data. Because of its low
count rate it is difficult to put an upper limit on the amplitude of
any inherent period. However, we consider it to be quite possible that
a period of 20 mins with an amplitude of, say, 10 percent could easily
go undetected.

We used the on-line tool {\tt PIMMS} (Mukai 1993) to convert the {\sl
ROSAT} HRI count-rate to flux assuming a Galactic column density of
$1\times10^{20}$ \pcmsq and an internal absorption of $5\times10^{20}$
\pcmsq (appropriate for an accreting binary source).  Assuming a
thermal bremsstrahlung emission spectrum of temperature of 20 keV a
count rate of 0.0091 ct/s gives an observed flux (0.2--2.4keV) of
3.6$\times10^{-13}$ \ergscm. This gives an observed X-ray
luminosity of 4.3$\times10^{29} d_{100}^{2}$ erg/s (0.2--2.4keV) and
an unabsorbed bolometric luminosity of 2$\times10^{30} d_{100}^{2}$
erg/s.

We can estimate the distance to RAT J1953+1859 using the absolute
magnitude in outburst verses orbital period relationship for hydrogen
(non-magnetic) accreting CVs (see Warner 1987 and Patterson
2009). Although there is some uncertainty in the peak brightness of
RAT J1953+1859 in outburst, we assume $V_{max}=$16.5 (cf \S
\ref{not}). To account for the apparent orientation of the disc we
assume a binary inclination $i=45^{\circ}$, which gives an inclination
corrected peak brightness of $V=$16.1 (cf Patterson 2009, eqn 2). For
an orbital period of 90 min, the $M_{V,max}, P_{orb}$ relationship
implies $M_{V}=$5.3 and therefore a distance of $\sim$1.5 kpc. Such a
distance would indicates an X-ray unabsorbed bolometric luminosity of
4.5$\times10^{32}$ erg/s. This implies that {\src} is one of the more
luminous dwarf novae observed using {\ros} (van Teeseling, Beuermann
\& Verbunt 1996).

van Teeseling \& Verbunt (1994) plot the X-ray Optical/UV flux ratios
for a number of CVs of different types. For sources not observed using
{\sl IUE} they use the relationship $log F_{(UV+Opt)}= -0.4m_{V} -
4.32$ as an approximate guide. For RAT J1953+1859 in a low state, this
gives $F_{UV+Opt}=3.3\times10^{-13}$ \ergscm. For an unabsorbed X-ray
flux over the energy range 0.1--2.0keV of $5.0\times10^{-13}$ \ergscm
this gives an $F_{X}/F_{UV}\sim$5. Although there is considerable
uncertainty in this calculation, this ratio suggests an orbital period
of approximately 1 hour (cf Figure 7 of van Teeseling \& Verbunt
1994).

\section{Discussion and Conclusion}

RAT J1953+1859 shows clear evidence for low and high accretion states
and has a clear X-ray counterpart. Further, it shows no emission from
He II (4686\AA) during quiescence, while the Balmer emission lines
appear rather broad (1800 km/s). During outburst, there is some
evidence that the light curve repeats on a period of $\sim$90
mins. This is consistent with the 80--90 min period seen in the V/R
ratio derived from spectra taken during quiescence.  All of these
characteristics point to {\src} being a CV and a dwarf nova in
particular.

Most dwarf novae with periods shorter than the period gap (2 hrs) are
SU UMa systems -- these dwarf novae experience super-outbursts which
are brighter and last much longer than normal outbursts. If we assume
for argument that the 90 min period is real, then this period would be
the signature of the super-hump modulation, which is typically a few
percent longer than the orbital period. The 80--90 min modulation seen
in the V/R ratio power spectra during quiescence is therefore likely
due to the binary orbital period.

What makes {\src} unusual is that it was discovered by means of high
amplitude QPOs seen during quiescence. QPOs have been seen in many
CVs, ranging from $\sim$100 sec to $\sim$2000 sec (see Warner 2004 for
a recent review of the whole range of quasi-periodic behaviour seen in
CVs). However, QPOs are mostly seen during a dwarf nova outburst, or
in the rise up to, or decline from an outburst. We know of only two
instances where quasi periodic behaviour has been seen in dwarf novae
in quiescence and both had periods which were shorter than that
observed in {\src}. V893 Sco showed oscillations on a period of 5.7
mins (and an amplitude of $\sim$0.2 mag) during one night of
observation (Bruch, Steiner \& Gneiding 2000) and WX Hyi which showed
QPOs with a period near 3 min (and amplitude of $\sim$0.1 mag) at
several quiescent epochs (Pretorius, Warner \& Woudt 2006).

G\"{a}nsicke (2005) gave a summary of the different means for
discovering CVs. While around half were discovered as a result of
their optical variability, all these systems were found as a result of
the source undergoing an outburst. {\src} is the first CV to be
identified as a result of high amplitude QPOs seen during quiescence.
We agree with Pretorius et al (2006) who note that the apparent lack
of QPO's found in dwarf nova in quiescence could be simply that people
have not looked for them. 

The RATS project has several million light curves in its
archive. We have so far concentrated on identifying sources which show
periodic behaviour in their light curves. A next step is to apply a
number of different variability measures. Since CVs show prominent
flickering behaviour (eg Bruch 1992) we expect to discover many more
CVs through this means in our survey. We therefore have good reason to
expect that high time resolution photometric surveys such as RATS are
likely to become an increasingly important tool for identifying
accreting binaries in general and cataclysmic variables in
particular. More complete samples of such systems may be selected by
this methods using, for instance, the LSST and VST.

\section{Acknowledgements}

Observations were made using the William Herschel Telescope, the Isaac
Newton Telescope and the Nordic Optical Telescope on La Palma. We
gratefully acknowledge the support of each of the observatories
staff. We thank Diana Hannikainen and Hanna Tokola for assisting with
the NOT observations. Some of the data presented here have been taken
using ALFOSC, which is owned by the Instituto de Astrofisica de
Andalucia (IAA) and operated at the Nordic Optical Telescope under
agreement between IAA and the NBIfAFG of the Astronomical Observatory
of Copenhagen. We thank the anonymous referee for some helpful
comments.

\end{document}